\documentclass[10pt,journal,compsoc]{IEEEtran}
\ifCLASSOPTIONcompsoc
  \usepackage[nocompress]{cite}
\else
  \usepackage{cite}
\fi

\ifCLASSINFOpdf
\else
\fi

\usepackage{cite}
\usepackage{amsmath,amssymb,amsfonts}
\usepackage{tabularx}
\usepackage{blindtext}
\usepackage{graphicx}
\usepackage{textcomp}
\usepackage{wrapfig}
\usepackage{multicol}
\usepackage{multirow}
\usepackage{color,colortbl}
\usepackage[normalem]{ulem}

\definecolor{mygray}{rgb}{0.86, 0.86, 0.86}
\newcommand{\hlgray}[1]{{\sethlcolor{mygray}\hl{#1}}}

\def\BibTeX{{\rm B\kern-.05em{\sc i\kern-.025em b}\kern-.08em
    T\kern-.1667em\lower.7ex\hbox{E}\kern-.125emX}}

\definecolor{abstractbg}{rgb}{0.89804,0.94510,0.83137}
\graphicspath{{imm/}}
\usepackage{makecell} %table cell on more lines
\usepackage{algorithm}
\usepackage{algpseudocode}
\usepackage{array,booktabs}
\usepackage{siunitx} % per tabelle piene di numeri
\usepackage{soul}
\usepackage{verbatim}
\usepackage{pifont}
\usepackage{footnote}
\makesavenoteenv{tabular}
\makesavenoteenv{table}
\usepackage[switch]{lineno}
\newcommand{\cmark}{\ding{51}}

\usepackage[printonlyused]{acronym}
\newcommand{\canc}[1]{\textcolor{red}{\st{#1}}}
\newcommand{\add}[1]{\textcolor{blue}{#1}}

\usepackage{hyperref}
\hypersetup{
     colorlinks = true,
     linkcolor = blue,
     anchorcolor = blue,
     citecolor = blue,
     filecolor = blue,
     urlcolor = blue
}

\begin{document}
\title{B-HAR: an open-source baseline framework for in-depth study of human activity recognition datasets and workflows}
\author{Florenc~Demrozi,~\IEEEmembership{Member,~IEEE,}
        Cristian~Turetta,~\IEEEmembership{Member,~IEEE,}
        and~Graziano~Pravadelli,~\IEEEmembership{Senior~Member,~IEEE}% <-this % stops a space
\IEEEcompsocitemizethanks{\IEEEcompsocthanksitem F. Demrozi is with the Faculty of Science and Technology Department of Electrical Engineering and Computer Science of the University of Stavanger, Norway.\protect\\
E-mail: florenc.demrozi@uis.no
\IEEEcompsocthanksitem C. Turetta and G. Pravadelli are with the Department of Computer Science of the University of Verona, Italy.\protect\\
E-mail: cristian.turetta@univr.it, graziano.pravadelli@univr.it}% <-this % stops an unwanted space
\thanks{Manuscript received November XX, 2022; revised XXX XX, 20XX.    \\ \textbf{This is the author’s draft of an article submitted to IEEE. Copyright may be transferred without notice, after which this version may no longer be accessible. A copyright notice will be added here upon submission, and additional information in the case of an acceptance/publication.}}}

\IEEEtitleabstractindextext{%
\begin{abstract}
Human Activity Recognition (HAR), based on machine and deep learning algorithms, is considered one of the most promising technologies to monitor professional and daily life activities for different categories of people (e.g., athletes, elderly, kids, employers) in order to provide a variety of services related, for example to well-being, empowering of technical performances, prevention of risky situation, and educational purposes.
However, at the state of the art, we lack a standard workflow that acts as a baseline for estimating the effectiveness and the efficiency of HAR methodologies, and thus evaluating their quality.
This makes the comparison among different approaches a challenging task. 
In addition, researchers can make mistakes that, when not detected, definitely affect the achieved results.
To mitigate such issues, this paper proposes an open-source, highly configurable framework, named B-HAR (Baseline-HAR), for the automatic definition and implementation of a baseline workflow that supports researchers in evaluating and comparing HAR approaches. 
B-HAR implements the most popular data processing methods for data preparation and the most commonly used machine and deep learning pattern recognition models. 
Moreover, B-HAR provides the capability to integrate user-defined machine/deep learning models by maintaining the data pre-processing steps unaltered.
B-HAR has been validated on 9 of the most famous HAR datasets containing data perceived by different sensor types (i.e., inertial, physiological, and environmental) positioned at different human body or environment points.
\end{abstract}

}

% make the title area
\maketitle

\IEEEdisplaynontitleabstractindextext

\IEEEpeerreviewmaketitle

% !TEX root = ../main.tex
\section{Introduction}\label{sec:intro}

\IEEEPARstart{I}{n} the last decade, with the advent of the Internet of Things (IoT), embedded sensors have started to be integrated into personal devices such as smartphones, smartwatches, clothes, and objects of daily life. 
This has pushed the definition of new research directions and the development of smart applications for Human Activity Recognition (HAR). 
Meanwhile, HAR has become a popular topic due to its importance in many areas, including health care, well-being, interactive gaming, sports, and monitoring of Activities of Daily Life (ADLs), in both controlled and uncontrolled settings~\cite{bianchi2019iot,poli2020impact}.%antunes2018survey 

Among the different application scenarios, HAR is becoming more and more relevant, in particular, for monitoring and coaching the older population, since, according to the 2019 World Population Prospects~\cite{demrozi2021exploiting,UN19}, in 2018, for the first time in human history, persons aged 65+ years outnumbered children under five years.
% They expect that in 2050 people aged 65+ years(1.5 billion) will outnumber adolescents and youth aged 15 to 24 years (1.3 billion). 
In fact, despite an increase in life expectancy, most people show a loss of self-efficacy as they age and a consequent reduction of the quality of their life~\cite{demrozi2019towards,demrozi2019indoor}.
This decrease in autonomy, in conjunction with the necessity of changing some habits and acquiring new behaviors (e.g., taking medicine at periodic intervals) concerning ADLs requires innovative use of new technologies/techniques such as HAR~\cite{demrozi2021exploiting}.
\begin{figure*}[t!]
\centering
\includegraphics[width=0.8\textwidth,page={2}]{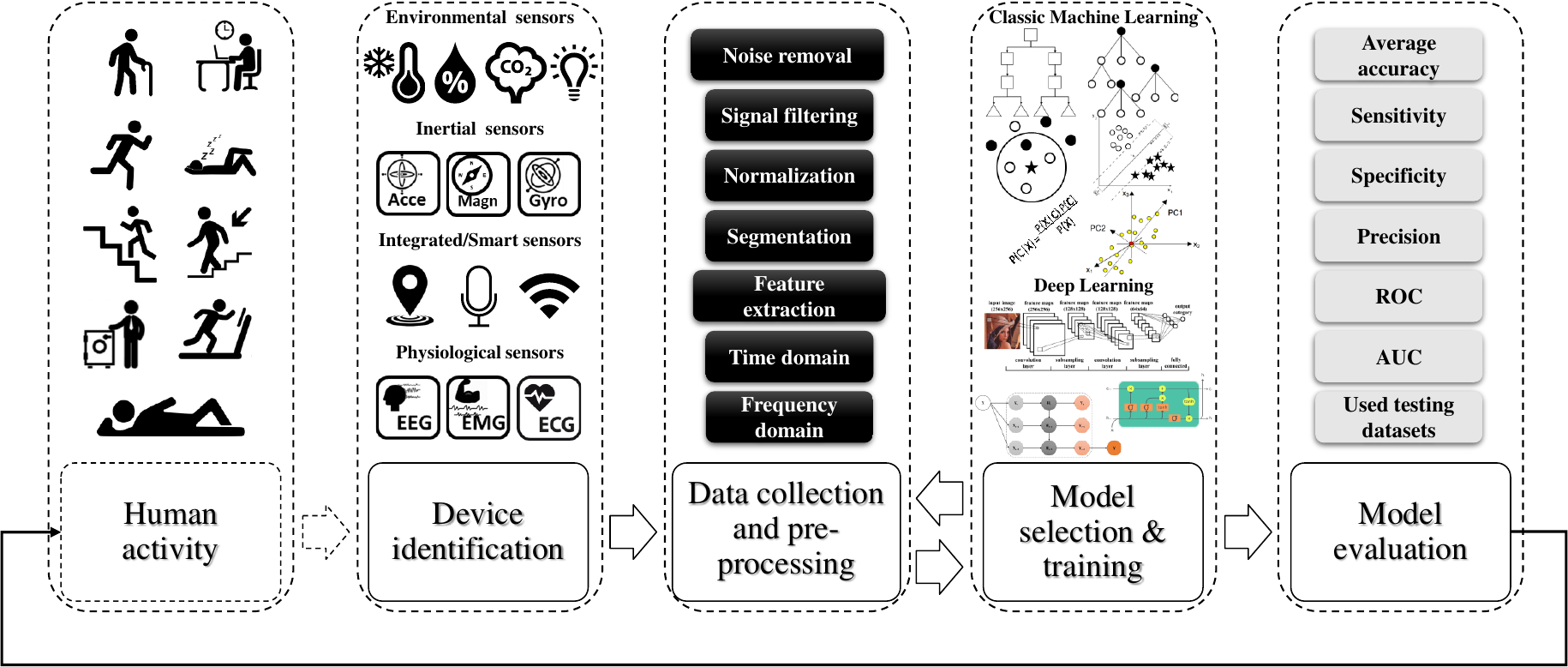}
\caption{Overview of the HAR workflow.}\label{fig:HARview}
\vspace{-0.5cm}
\end{figure*}
Despite its increasing importance, research on HAR encounters multiple difficulties. 
Among them, the most critical are: i) the variety of daily activities to be recognized, ii) the intra-subject and inter-subject movement variability, iii) the trade-off between performance and privacy, iv) the need of computational efficiency and their actual availability in embedded and portable devices where HAR algorithms should run, and v) time-consuming data annotation processes~\cite{lara2012survey,fullerton2017recognizing,demrozi2021inertial}.

Concerning the data elaborated by HAR algorithms, they are typically collected from cameras and ambient sensors (e.g., temperature, humidity, brightness, seismic) as well as wearable embedded sensors (e.g., inertial or physiological sensors) ~\cite{pucci2020human,gambi2020adl,lu2020efficient}. %,liu2016action,zeng2014convolutional,jain2017human}.
Cameras are broadly used in HAR, nevertheless, collecting video data presents privacy issues, and in addition, their elaboration requires high computational resources. 
For these reasons, many researchers prefer to work with ambient and embedded sensors~\cite{donahue2015long,burgos2020ear}.
In particular, inertial sensors have shown excellent results in HAR, and their use in combination with other types of sensors is rising rapidly. 
The diffusion of such sensors is strongly related to their capacity to map the human body movement directly~\cite{demrozi2020human} and also human psychological state (e.g., pain level~\cite{demrozi2020pain}). 
Besides, inertial sensors present affordable costs and they can be integrated into most of wearable objects people own.
Nevertheless, with modern improvements in sensor networks research, we are on the path of revolutionary low-cost monitoring systems embedded within our homes and daily life environments~\cite{black2004pervasive,kameas2010pervasive}.

Concerning the algorithmic design, HAR has seen an increasing interest in Deep Learning (DL) methods, 
%\footnote{\add{imposing itself as the new state-of-the-art HAR standard}},
mainly due to their capabilities to work with raw data that, in principle, do not require explicit data preprocessing techniques as normalization, filtering, or feature extractions~\cite{kanjo2019deep}. %zeng2014convolutional 
However, DL models requires large human activity  datasets to produce. The lack of such datasets requires to apply different processing techniques~\cite{demrozi2020human}. 
Another prominent characteristic of DL models that affects their usability is related to the identification of their optimal architecture.
On the other side, in classic Machine Learning (ML) models, the application of grid search techniques reduces the effort that users need to apply for identifying the most suitable recognition model. 
Instead, in DL, the application of grid search techniques provides only the possibility to dynamically set the model hyperparameters (i.e., batch size, training epochs, optimization algorithms, learning rate, momentum, weight initialization, activation functions, and dropout regularization) and not the architecture (i.e., number of neurons and hidden layers, convolutional layers, pooling layers, etc.) of the DL model~\cite{brownlee2016grid}.
%In this scenario, HAR becomes one of the most promising solutions to assist older people’s daily life and one of the most used technologies in health care delivery~\cite{wang2019deep}. 

Moreover, the design of a workflow is a challenging and error-prone activity that may definitely affects the accuracy of the HAR methodology.
Figure~\ref{fig:HARview} presents the typical workflow for
HAR \add{algorithmic design}. It generally consists of: 1) the identification of data sources, i.e., sensors and devices used to collect data, 2) the data collection and preprocessing, 3) the model selection and training, and 4) the evaluation of the achieved results. 
Despite this general workflow, the data preprocessing and the model selection phases include a high number of non-standardized steps, which represents the critical aspects for the quality of the HAR approach. 
In particular, the data preprocessing includes different elaborations, such as normalization, noise removal, balancing, or feature extraction. 
Their application profoundly impacts the recognition model performance. 
Indeed, no well-defined workflow that identifies how these data processing steps should be applied exists, as well as there are no precise rules for the model selection and training.

The lack of such well-defined workflow makes it very difficult to analyze, compare and evaluate the quality of different HAR approaches, even by adopting the same dataset. 
Some of the most common errors during the implementation of this workflow concern:
\begin{itemize}
    \item data errors (i.e., null values, infinite values, or corrupted values) are not handled;
    \item normalization is applied to the initial datasets instead of applying it to the training dataset and then to the validation and testing dataset;
    \item incorrect training and testing (train/test split) procedure;
    \item performance evaluation of multi-class problems is provided only in terms of accuracy, instead of using metrics such as sensitivity, specificity, precision, and F1-Sore. 
\end{itemize}
Such errors are mainly due to the fact that the workflow steps are usually applied separately, and there is no tool that integrates them, making their use transparent to the user.

In addition, many datasets which represent human activities, perceived through inertial sensors (i.e., accelerometers, gyroscopes, magnetometers), physiological sensors (i.e., electromyography (EMG), electrocardiogram (ECG), or electroencephalogram (EEG)) are available in the literature~\cite{kwapisz2011activity,roggen2010collecting,bachlin2009wearable,reiss2012introducing,zappi2007activity,anguita2013public, zhang2012usc, stisen2015smart, banos2014mhealthdroid, banos2012benchmark}. 
%Among them the most known and used are WISDM v1~\cite{kwapisz2011activity}, WISDM v2~\cite{kwapisz2011activity}, Opportunity~\cite{roggen2010collecting}, DAPHNET~\cite{bachlin2009wearable}, PAPAM~\cite{reiss2012introducing}, Skoda~\cite{zappi2007activity}, UCI-HAR~\cite{anguita2013public}, USC-HAD~\cite{zhang2012usc},   HHAR~\cite{stisen2015smart}, mHealth~\cite{banos2014mhealthdroid}, and REALDISP~\cite{banos2012benchmark}.
However, to the best of our knowledge, at the state-of-the-art, there is no work (article or tool) that can be used as a baseline for fairly comparing different HAR methodologies.

To fill in the gap, this paper propose the Baseline-HAR (B-HAR)\footnote{\href{https://github.com/B-HAR-HumanActivityRecognition/B-HAR}{B-HAR github page.}} Python framework, as an open-source tool for simplifying the set up and the configuration of HAR workflows. 
%
% \canc{that, starting from the target input dataset\footnote{Represented by time series of data perceived by inertial, physiological sensors and/or other type of data.}, provides the user with: i) the possibility of easily configuring the different phases of data preprocessing and training, thus avoiding errors; ii) efficiency indicators of the most used classification models applied to the target dataset}.
%
In particular, the main characteristics provided by B-HAR are:
\begin{itemize}
\item unification of the main workflow of Figure~\ref{fig:HARview} into a single framework, making the application of the various preprocessing steps transparent to the user;
\item definition of a precise order of application of the data preprocessing steps, to minimize errors;
\item provision to the users baseline information about the most famous HAR datasets~\cite{kwapisz2011activity,lockhart2011design,bachlin2009wearable,stisen2015smart,reiss2012introducing,banos2014mhealthdroid}, %~\cite{kwapisz2011activity,roggen2010collecting,bachlin2009wearable,reiss2012introducing,zappi2007activity,anguita2013public, zhang2012usc, stisen2015smart, banos2014mhealthdroid, banos2012benchmark}
such that new HAR approaches can be exhaustively compared with respect to existing results;
\item given a new HAR dataset, automatic application of the grid search technique on the most used pattern recognition models to provide the user with  baseline results;
\item possibility for the users of implementing their own HAR (ML or DL-based) models and of integrating them in B-HAR, allowing the reuse of previously-implemented preprocessing workflows, tested on other models or datasets.
\end{itemize}

We finally emphasize that by presenting B-HAR, this paper is not intended to propose a new HAR model or HAR standard workflow. 
Our goal is, instead, to provide the scientific community with a unified, but customizable, framework for enabling error-free comparisons among existing and new recognition models.

The rest of the paper is organized as follows. 
%Section~\ref{sec:back} introduces some preliminary concepts. 
Section~\ref{sec:over} describes B-HAR and the underpinning methodology. 
Section~\ref{sec:res} presents the results of an exhaustive experimental campaign.
Subsequently, Section~\ref{sec:dis} briefly discusses current limitations of the proposed approach and future work. 
%Section~\ref{sec:rel} gives an overview of the related works and.
Finally, Section~\ref{sec:conc} concludes the papers with some remarks.
\begin{figure*}[t!]
\centering
\includegraphics[width=0.8\textwidth,page={3}]{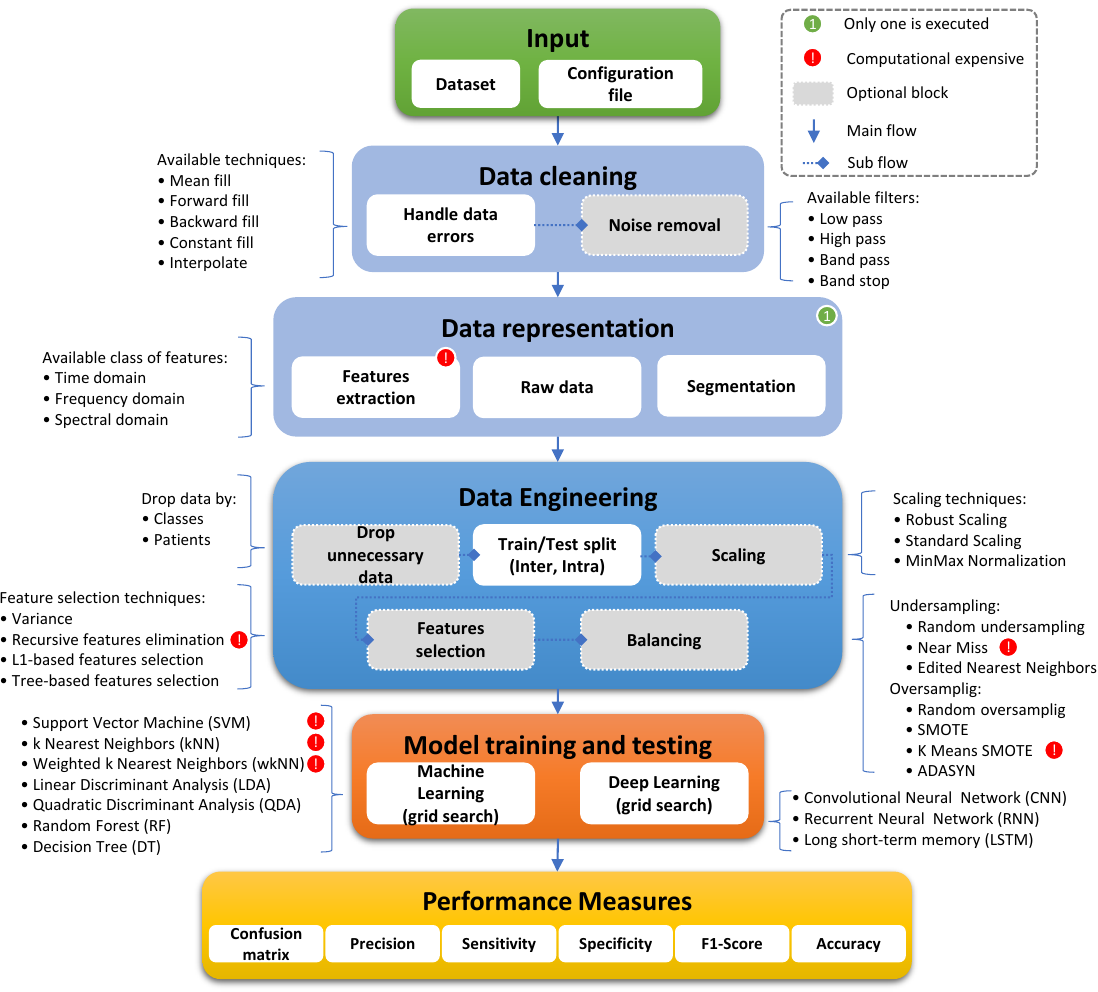}
\caption{Overview of the B-HAR structure.}\label{fig:b-har_structure}
\end{figure*}
% \input{sec/back.tex}
% !TEX root = ../main.tex
%----------------------------------------------------
\section{Methodology}\label{sec:over}
B-HAR comprises six dedicated modules: one input module, one output module, and four computation modules including various sub-modules.
Its structure is shown in Figure~\ref{fig:b-har_structure}. 
Red bullets identify computationally heavy and time-consuming operations. 
The green bullet refers to a computation module that executes only one of its sub-modules. 
Grey sub-modules identify optional computations that B-HAR users may use or not.
Internal sub-modules execute in the order indicated by the dashed lines.
Details on the structure are reported in the following sections.
%----------------------------------------------------
% \subsection{Input module}
\subsection{Input}
In order to start the analysis, the first step is to perform the B-HAR object instantiation.
It takes several input parameters, further detailed in this section.
In particular, the dataset input path can be a single file or a directory. 
In the case of a directory, B-HAR will merge all files inside it to create the entire dataset, i.e., values of the signals perceived by sensors during human activities, enriched by information concerning the testing subject identity, the data collection session, and the performed activity at a specific timestamp.

% B-HAR takes in input two files containing, respectively, i) the dataset, i.e., values of the signals perceived by sensors during human activities, enriched by information concerning the testing subject identity, the data collection session, and the performed activity at a specific timestamp, and ii) a configuration file that defines the library workflow.

%----------------------------------------------------
\subsubsection{Dataset format}
Since HAR datasets present different structures and information types, B-HAR can flexibly handle different dataset formats. 
Table~\ref{tab:collected_data} shows the generic, customizable, structure of the input dataset that can be provided to B-HAR; \cmark identifies mandatory data, while [\cmark] identifies optional data.
B-HAR requires a dataset composed of signals from one or more sensors ($Sensor_{xyx}^i$), a label ($A$) indicating the human activity performed by the testing subject, and his/her identifier ($T$). 
% \canc{B-HAR can handle datasets including data from different types of sensors (e.g., accelerometers, gyroscopes, magnetometers, or other physiological sensors) that can be related to different activities, each one identified by a different session identifier ($S$).}
In general, these data are collected during supervised collection sessions ($S$) where subjects ($T$) are instructed to perform only one activity (e.g., walking) or a series of consecutive activities (e.g., sitting, standing up, standing, walking, standing, sitting down, sitting).
B-HAR handles both types of data collection approaches (i.e., one activity/one session, many activities/one session).
% $S$ identifies the data collection session identifier. 
\begin{table}[ht!]
\centering
\resizebox{0.48\textwidth}{!}{
\begin{tabular}{c|ccc|ccc}
\cline{2-7}
& \multicolumn{3}{c|}{Data} & \multicolumn{3}{c}{Info}\\
\hline
$Time (Ti)$         &
$Sensor_{xyz}^1$    &
$...$               &
$Sensor_{xyz}^n$    &
$A$                 &
$T$                 &
$S$                 \\
\hline
\hline
[\cmark]            &
\cmark              &
....                &
[\cmark]            &
\cmark              &
\cmark              &
[\cmark]            \\
\hline
\hline
\multicolumn{1}{c}{$n~~\geq~1$}                     &
\multicolumn{3}{c}{\cmark~$\equiv~is~mandatory$}       & \multicolumn{3}{c}{[\cmark]~$\equiv~is~optional$}   \\
\multicolumn{1}{c}{A = Activity}                    & 
\multicolumn{3}{c}{T = Testing subject}             & 
\multicolumn{3}{c}{S = Session}                     \\
% \multicolumn{7}{c}{
% A = Activity, T = Subject, S = Session
% } \\
\end{tabular}
}\vspace{0.1cm}
\caption{B-HAR dataset structure.}\label{tab:collected_data}
%\vspace{-0.5cm}
\end{table}~~\\
Moreover, our framework can handle input datasets were data appears in different orders as 
\begin{itemize}
    \item $Ti, Data, A, T, S$;
    \item $Data, A, T, S, Ti$;   
    \item $A, T, S, Data, Ti$;
    \item $A, T, S, Ti, Data$;
    \item $S, A, T, Ti, Data$;
    \item and other configurations. 
\end{itemize}

% Besides, if time information lacks, B-HAR users have to indicate, into the configuration file, the discrete sampling frequency adopted by the sensors.
Besides, if time ($Ti$) information lacks, B-HAR uses the discrete sampling frequency adopted by the sensors. 
From now on, all the processing steps are performed by taking into account the applied data collection methodology (i.e., one activity/one session, or many activities/one session). 
In the many activities/one session approach, B-HAR provides the possibility to generate transition classes that generally are not handled by the users (e.g., a sequence of activities as [sitting, standing, walking, standing up, walking] is represented as [sitting, sitting\_to\_standing, standing, standing\_to\_walking, walking, walking\_to\_standing, standing, standing\_to\_walking, walking ]).

%----------------------------------------------------
\subsubsection{Configuration parameters}
The B-HAR object provides the user with total control over the computation modules and relative sub-modules shown in Figure~\ref{fig:b-har_structure}. 
Through its initialization parameters, B-HAR users can easily modify the data processing workflow and test different configuration pipelines. 
Table~\ref{tab:config_parameters} shows the implemented configuration parameters (Column 2) alongside with their description (Column 3) and possible usable values (Column 4).
% \subsubsection{Configuration file}
% The B-HAR configuration file provides the user with total control over the computation modules and relative sub-modules shown in Figure~\ref{fig:b-har_structure}. 
% Through such a file, B-HAR users can easily modify the data processing workflow and test different configuration pipelines. 
% Table~\ref{tab:config_parameters} shows the implemented configuration parameters (Column 2) alongside with their description (Column 3) and possible usable values (Column 4).

\begin{table*}[!t]
\centering
\begin{tabular}{c| p{3.5cm} | p{6.5cm} | p{5cm}}%{\textwidth}{c|c|c|c}%
\hline
Nr. & Parameter & Description & Values Range \\
\hline
\hline % dataset 
 1 & \texttt{dataset\_path}                 & Path to the dataset directory.                                            & String.  \\
 6 & \texttt{dataset\_sampling\_freq}       & Sampling frequency of loaded dataset in Hz.                               & Integer \\
 4 & \texttt{header\_format}                & Header format of the input dataset.                                       & \emph{TiDA}, \emph{TiDAT}, \emph{DA}, \emph{DATS}, \emph{etc.}\\
 3 & \texttt{derive\_header}                & The input dataset file contains also the header or should it be derived   & Bool.\\
 2 & \texttt{separator}                     & .csv separator                                                            & Separator character \\
 2 & \texttt{dataset\_type}                 & the type of data in the dataset                                           & \emph{continous}, \emph{fragmented}\\
 5 & \texttt{segment\_time}                   & Time for desired window length, in seconds.                               & Float \\
 
 7 & \texttt{overlap} & Overlap between time windows, in seconds. & Float \\
 8 & \texttt{group\_by} & Show stats by attribute. & \emph{CLASS}, \emph{P\_ID} \\
% 9 & \texttt{resampling} & For unbalanced dataset. & \emph{under}, \emph{over}, \emph{standard} \\
\rowcolor[gray]{0.8} 9 & \texttt{representation} & Data treatment type. & \emph{\textbf{\uline{segmentation}}}, \emph{raw}, \emph{\textbf{\uline{features\_extraction}}} \\
\rowcolor[gray]{0.8} 10 & \texttt{features\_domain} & Available domains for features extraction .& \emph{statistical}, \emph{spectral}, \emph{temporal}, \emph{\textbf{\uline{all}}}.\\
\rowcolor[gray]{0.8} 11 & \texttt{features\_selection} & Toggle features selection. & Bool (\emph{\textbf{\uline{true}},~false}) \\
\rowcolor[gray]{0.8} 12 & \texttt{models} & Implemented ML and DL models. & \textit{\textbf{\uline{kNN}},~wkNN,~\textbf{\uline{LDA}},~\textbf{\uline{QDA}},~SVM,~\textbf{\uline{RF}},~\textbf{\uline{DT}}, \textbf{\uline{CNN}}, NN, LSTM, RNN} \\
13 & \texttt{user\_models} & User defined ML or/and DL models. & \textit{\textbf{{Must be implemented with sklearn or tensorflow}}} \\

 % preprocessing

\rowcolor[gray]{0.8} 14 & \texttt{scaling\_method} & Normalize data. & \emph{none}, \emph{minmax}, \emph{\textbf{\uline{robust}}}, \emph{standard} \\
\rowcolor[gray]{0.8} 15 & \texttt{split\_method} & Train/Test split method. & \emph{\textbf{\uline{intra}}}, \emph{inter} \\
\rowcolor[gray]{0.8} 16 & \texttt{feature\_selection\_mtd} & Features selection technique. & \emph{variance}, \emph{l1}, \emph{\textbf{\uline{tree-based}}}, \emph{recursive} \\
17 & \texttt{n\_features\_to\_selec}t & Number of features to select, available only with \emph{recursive} selection method. & Integer \\
18 & \texttt{balancing\_method} & For unbalanced dataset, balancing techniques. 

(Under $\bullet$ Over sampling techniques)& 
\emph{random\_under}, \emph{near\_miss}, \emph{edited\_nn} $\bullet$

\emph{random\_over}, \emph{smote}, \emph{adasyn}, \emph{kmeans\_smote}.\\
%19 & \texttt{data\_balancing\_method} & For unbalanced dataset, balancing techniques. & under $\in${\emph{random}, %\emph{near\_miss}, \emph{edited\_nn}}
%over $\in${\emph{random}, \emph{smote}, \emph{adasyn}, \emph{kmeans\_smote}, \emph{random\_over}}.\\

 % cleaning

\rowcolor[gray]{0.8} 19 & \texttt{replace\_error\_method} & Substitute NaN and Inf values method.  &  \emph{\textbf{\uline{mean}}}, \emph{forward}, \emph{backward}, \emph{constant}, \emph{interpolate} \\
20 & \texttt{constant\_value} & The value which substitute NaN and $\pm\infty$ values. & Float \\
\rowcolor[gray]{0.8} 21 & \texttt{filter\_type} & Data filtering techniques. & \emph{\textbf{\uline{lowpass}}}, \emph{highpass}, \emph{bandpass}, \emph{bandstop} \\
\rowcolor[gray]{0.8} 22 & \texttt{filter\_order} & The order of the applied  filter. & Integer (\textbf{\uline{e.g., 4}}) \\ 
\rowcolor[gray]{0.8} 23 & \texttt{filter\_cutoff} & Low/High-Cutoffs frequency in Hz. & Integer (\textbf{\uline{e.g., 20 Hz}}).\\
%24 & \texttt{low\_cut} & Cutoff frequency in Hz. & Float.\\
%25 & \texttt{high\_cut} & Cutoff frequency in Hz. & Float \\

 % training

\rowcolor[gray]{0.8} 24 & \texttt{test\_size} & The size of test data. & Float (\textbf{\uline{e.g., 25\%}}) \\
\rowcolor[gray]{0.8} 25 & \canc{\texttt{epoch}s} & Training epochs for CNN & Integer (\textbf{\uline{e.g., 100}}) \\
\rowcolor[gray]{0.8} 26 & \canc{\texttt{k\_fold}} & Number of fold for CNN \emph{k-fold} training & Integer (\textbf{\uline{e.g., 3}}) \\
\rowcolor[gray]{0.8} 27 & \canc{\texttt{loss\_threshold}} & Bias used in CNN model ensembles & Float (\textbf{\uline{e.g., 0.4}}) \\
\rowcolor[gray]{0.8} 28 & \texttt{use\_features} & If features selection is enabled, extracted features are used as input of the CNN. & Bool (\textit{\textbf{\uline{true}},~false})\\

\hline
\hline
\end{tabular}\vspace{0.1cm}
\caption{B-HAR configuration parameters 
\scriptsize{
(\textbf{{\hlgray{gray lines}}} defines the configuration used in the experimental results section).
}}\label{tab:config_parameters}
\vspace{-0.5cm}
\end{table*}
\subsection{Data cleaning}
The quality of the collected data depends mainly on two factors: i) environment and hardware noise, and ii) data collection technology and architecture, which often leads to the loss or corruption of the data during its transmission.
Missing or corrupted data must be examined and integrated to maintain the structure and the information of the time series. 
Instead, environmental and hardware noise is handled through a noise removal phase.
The data cleaning module is devoted to handling such issues.
% \add{However, since, as early mentioned data collection is performed generally in two different modalities: i) one activity/one session, and ii) many activities/one session, such phases are anticipated by the addition of a column that }

\subsubsection{Handling of data errors}
Missing and inconsistent data, identified as NaN, or Inf, represent one of the main obstacles for an accurate analysis of observed data.
B-HAR provides five different methods to handle missing data:
\begin{itemize}
\item \textit{bfill}: it is used to backward fill the missing values in the time series;
\item \textit{ffill}: it is used to forward fill the missing values in the time series;
\item \textit{interpolate}: it performs linear interpolation for missing data based on previous and successive data points;
\item \textit{constant(k)}: it substitutes missing values with a constant value $k$;
\item \textit{mean}: it substitutes missing data with the arithmetic mean of the previous and successive observations.
\end{itemize}

B-HAR returns an error if the input dataset has more than $M\%$\footnote{$M\%$ of the samples in a time series column. By default $M=5$} ($M$ value is provided as input by the B-HAR user) of not adjacent missing values~\cite{pratama2016review}. 

Otherwise, if the $M\%$ of missing values are related to a specific session or activity (intended as a continuous time series of one activity), the user has the possibility to drop them.
%the data related to the session or activity containing an high percentage of missing data. 
% \canc{the user configures B-HAR,  such that this data collection session or activity is excluded from the dataset.}

\subsubsection{Noise removal}
Time series noise removal is an essential and indisputable step in HAR signal processing. 
B-HAR integrates four different types of filters~\cite{sensormotion}:
\begin{itemize}
\item \textit{low-pass filters} pass signals with a frequency lower than a selected cutoff and attenuate the rest; 

\item \textit{high-pass filters} pass signals with a frequency higher than a selected cutoff and attenuate the rest;

\item \textit{band-pass filters} pass signals within a specific range and attenuates frequencies outside it;

\item \textit{band-stop filters} pass most frequencies unaltered, but attenuate those in a specific range.
\end{itemize}

%The high, low-pass, and band-stop filters take as input the cat-off frequency, the input signal's sampling rate, and the filter order. 
%Instead, the band-pass filter takes as input the same parameters of the previous filters except for the cutoff frequency provided as a couple of values instead of a single value.
B-HAR users can also choose not to apply noise removal techniques, relying entirely on the capabilities of pattern recognition models.\\

\noindent The data handling and noise removal modules are not applied to all the datasets but are applied to a single data collection session or activity based on the dataset type\footnote{one activity/one session, or many activities/one session}.
%----------------------------------------------------
\subsection{Data representation}
In general, the collected data can be represented in different forms: \begin{itemize}
    \item raw data with no transformation;
    \item after features extraction techniques have been applied;
    \item after segmentation (i.e., grouping them in time windows).
\end{itemize} 

Since single data observations in a time series are not statistically informative, segmentation or features extraction is generally applied. 
Moreover, HAR models resonate with ongoing activities over time and not on single observations. 
With such an aim, segmentation (especially in deep learning) and feature extractions (especially in classical machine learning) have become state-of-the-art procedures for HAR.

B-HAR allows users to select one of the aforementioned data treatment procedures. 

\subsubsection{Raw data}
In this representation form, B-HAR does not perform any treatment, and the pre-processing module takes in input the dataset generated by the cleaning module. 
This representation becomes valid when the user has already prepared the data offline and he/she wants to use B-HAR to study different HAR models.

\subsubsection{Segmentation}
Given a %starting from an input time series sampled at a fixed frequency (e.g., 100 HZ) creates an unique time windows of 
$n \times 1$ time series dataset $HAR_D$, sampled at a fixed frequency $S_f$ (expressed in Hertz), and a time-window segment of dimension $T_w$ (expressed in seconds), the basic segmentation process implemented in B-HAR
% \begin{itemize}
%     \item time series $|HAR_D| = n \times 1 $, sampled at a fixed frequency $S_f~(Hz)$,
%     \item time window segment dimension $T_w~(seconds)$.
% \end{itemize}
returns a dataset $SHAR_D$ of dimension $(n/(T_w*S_f)) \times (T_w*S_f)$.

For example, given a $100 \times 1$ time series dataset $HAR_D$, a sampling frequency $S_f = 25$Hz, and a time window $T_w = 2$ seconds, the segmentation process returns a dataset $SHAR_D$ of dimension $ (100/(25*2)) \times (2*25)$, i.e.,  $2 \times 50$.
Nevertheless, many methodologies maintain an overlapping fragment between two consecutive segments. Such a fragment provides the model with information on the preceding segment context. 
B-HAR can also handle this overlapping if specified by the user.

%\noindent In conclusion, the preprocessing module will take in input the $SHAR_D$ dataset.
\subsubsection{Features extraction}
The feature extraction process explores the time and frequency domains of the raw data.
Time-domain features are mostly used because their extraction process requires a lower computation effort than the extraction of frequency-domain features.~\cite{barandas2020tsfel,shoaib2015survey}.
%Furthermore, since standard pattern recognition models are not suitable for raw data, this phase is anticipated by a segmentation step during which time-series sensor data is segmented before extracting features. 
%Table~\ref{tab:tf_features} shows the most commonly used time and frequency domain features. Such features are usually combined, further increasing the recognition accuracy.

B-HAR, given the original time series dataset, the type of features to be extracted in the desired domain (time and/or frequency), the time-window size, and the overlapping size, returns the corresponding time and/or frequency feature-based representation.
B-HAR uses the \href{https://github.com/fraunhoferportugal/tsfel}{Time Series Feature Extraction Library (TSFEL)} presented in~\cite{barandas2020tsfel}.
%B-HAR feature extraction module takes in input five parameters, i) HAR dataset produced by the data representation module, ii) the sampling frequency, iii) the time window size, iv) the overlap size, and the v) type of feature to extract. 
%----------------------------------------------------
Feature extraction is of primary importance if the sensors sample at a non-fixed frequency; B-HAR provides the possibility to create windows with a fixed time dimension (e.g., 2 seconds) regardless of the system sampling frequency. For example, if we have a time window of 2 seconds composed of 127 samples and another time window of 2 seconds of 100 samples, after the features extraction steps, these time windows will be represented by  150 features each.

\subsection{Data Engineering} % Da definire
This module adopts a set of techniques to reduce the dataset size by dropping unnecessary data, selecting a subset of extracted features, and balancing the dataset. 
In addition, it standardizes the data through scaling approaches.

\subsubsection{Drop of unnecessary data}
One of the main issues in HAR, and pattern recognition techniques in general, is represented by the presence in the dataset of imbalanced class distribution and noisy data related to a specific tester (subject), activity (class), or session.
To solve such an issue, B-HAR provides the possibility of excluding data from the input dataset.
In particular, B-HAR can exclude data related to:
\begin{itemize}
\item a single subject or a group of subject;
\item a single activity or a group of activities;
\item a single data collection session or a group of sessions.
\end{itemize}

\subsubsection{Training/testing approach}
In pattern recogniton, data usually is split into two (training and testing) or three (training, validation, and testing) subsets.
The recognition model is then trained on the training dataset (and cross-checked with the validation dataset when required) to make predictions on the testing dataset.
However, in HAR, train and test datasets are usually partitioned by considering the presence of data collected from several different testing subjects~\cite{bishop2006pattern}.
For such a purpose, B-HAR presents three different types of train/test splitting approaches:

\begin{itemize}
\item \textit{Inter-subjects}: % tra gruppi diversi
The inter-subject train/test partitioning approach works as follows: B-HAR users select a specific subset of subjects creating the testing dataset (e.g., 10 out of 100 total subjects), and the remaining subjects represent the training dataset (e.g., 90 out of 100 total subjects). 
Using this approach, we are sure that there is no overlap between the training and test datasets, minimizing the possibility of overfitting and in-depth testing of the model's generalizability capabilities. 

\item \textit{Intra-subjects}: % al interno di un gruppo
The intra-subject train/test partitioning approach works as follows: B-HAR users perform a traditional hold-out over the activity dataset. 
In particular, the datasets is initially divided into training and testing dataset (e.g., 75\% train and 25\% test). Subsequently, a classic $k$-fold cross-validation approach is applied only over the training dataset.
Finally, the trained models is tested over the testing dataset. 

\item \textit{Inter-sessions}: % al interno di un gruppo
In the intra-session train/test approach, B-HAR uses part of the performed data collection sessions of specific subjects as a training set and the remanent sessions as the testing set. 
However, this type of testing approach requires more sessions for each subject.
\end{itemize}
\noindent When dealing with HAR data, the best testing methodologies are inter-subjects or inter-session.

\subsubsection{Scaling}
During the training phase, features with higher values govern the training process. However, such features do not primarily represent the dataset's characteristics or the final accuracy of the pattern recognition model. 
Data scaling transforms multi-scaled data to the same scale, where all variables positively influence the model, thus improving the stability and the performance of the learning algorithm~\cite{bishop2006pattern}.
Besides, when working with datasets where different features represent every single sample, the datasets perform independent scaling for every single feature.
B-HAR provides the following scaling techniques:
\begin{itemize}
\item \textit{Robust scaling}:
it scales each feature of the dataset $HAR_D$ by subtracting the median ($HAR_D^{Q_2}$) and then dividing by the Interquartile Range (IQR). The IQR is defined as the difference between the third and the first quartile  ($HAR_D^{Q_3}) - HAR_D^{Q_1}$). The robust scaler of a dataset $HAR_D$ is expressed as:
\begin{equation}
HAR_D^{norm} = \frac{HAR_D-HAR_D^{Q_2}}{HAR_D^{Q_3} - HAR_D^{Q_1}}
\end{equation}
This scaler uses statistics that are robust to outliers, in contrast with the other scalers, which use statistics that are highly affected by outliers such as the maximum, the minimum, the mean, and the standard deviation.

\item \textit{Standard scaling (z-Score)}:
it maps the data into a distribution with mean 0 and a standard deviation 1. 
Each normalized value is computed by subtracting the corresponding feature's mean and then dividing by the standard deviation.
The standard scaler of a dataset $HAR_D$ is expressed as:
\begin{equation}
HAR_D^{norm} = \frac{HAR_D - HAR_D^u}{HAR_D^s}
\end{equation}
where $HAR_D^u$ is the mean of the training dataset, and $HAR_D^s$ is the standard deviation of the training datasets.

\item \textit{Min-Max normalization}:
it rescales the feature to a fixed range by subtracting the minimum value of the feature and then dividing by the range:
\begin{equation}
HAR_D^{norm} = \frac{HAR_D - HAR_D^{min}}{HAR_D^{max} - HAR_D^{min}}
\end{equation}
\end{itemize}
In particular, a normalizer is fitted with the training dataset, and then the same normalizer is used to normalize the testing data. However, a common error that researchers of the topic commit if they are new to the ML/DL field is to perform scaling before the train/test split step over the dataset.

\subsubsection{Features selection}
A higher number of features do not necessarily imply better results from the pattern recognition model. 
This is because features can positively or negatively impact the recognition process. 
For this purpose, feature selection techniques identify and order the features by importance. 
Starting from a training dataset, such techniques automatically select features that contribute most to the prediction accuracy, reducing the model's dependence on irrelevant features. 
Moreover, eliminating features reduces the time required for the training and testing phases.
The main benefits of feature selection techniques are i) reduction of the overfitting by eliminating redundant data that consequently reduces also noise-related errors, ii) improvement of the accuracy since misleading data are eliminated, and iii) reduction of the training time due to fewer data points~\cite{chandrashekar2014survey,bishop2006pattern}.\\

B-HAR integrates the following feature selection techniques:
\begin{itemize}
\item \textit{Variance}:
It removes all features whose variance is lower than a defined threshold. By default, it removes zero-variance features, i.e., features with the same value in all samples.

\item \textit{Recursive Features Elimination (RFE)}: 
It fits a model and removes the less critical feature (or features) until a defined number of features is reached, without knowing how many features are valid.
Features are ranked based on their importance and by recursively eliminating a small number of features per loop.  
Besides, RFE attempts to eliminate dependencies and collinearity that may exist in the model. 
Moreover, to determine the optimal number of features, k-fold cross-validation is used with RFE to score different feature subsets and select the best scoring collection.

\item \textit{Lasso regularization (L1)}:
%As already introduced, feature selection techniques aim to remove unimportant features based on a given importance threshold. 
%Apart from specifying the threshold numerically, there are built-in heuristics for finding a threshold. 
The Lasso regularization linear model estimates sparse coefficients starting from the non-zero coefficients returned by a linear regression model, thus effectively reducing the number of features the given solution is dependent upon. 

\item \textit{Tree-based}:
It is used to compute impurity-based feature importance, which can be used to discard irrelevant features in cooperation with other feature selection techniques. 
Tree-based estimators, by definition, internally create an ordering of the features representing the training dataset, which makes them very suitable as feature selection methods.
\end{itemize}

\subsubsection{Balancing}
%Imbalanced data typically refers to a classification problem where the number of observations of a particular dataset is not equally distributed (e.g., many observations for a particular class and a smaller number for one or more other classes. 
Training a HAR model on an imbalanced dataset can introduce unique challenges during the model training process, returning an influenced recognition model, especially when the contained data are not linearly separable. 
However, there are two categories of techniques that manage this issue at the state-of-art: i) oversampling the less populated classes to the same number of observations of the most populated class, and ii) undersampling the most populated classes to the same number of observations of the less populated class~\cite{imbalance}.
B-HAR provides different techniques to handle such issues:

\begin{itemize}
\item \textit{Undersampling random}:
It
%starting from classes with more observations than the less populated class of the dataset,
randomly eliminates a subset of observations of the must populated classes such that all classes have the same population (equal to the initial less populated class).
The main limitation is that the removed observations can be more informative than the kept ones.

\item \textit{Undersampling near miss}:
%randomly eliminates observations from the most populated classes. 
Given two observations, which are part of two different classes and are very similar to each other, it eliminates the observation belonging to the most populated class.

\item \textit{Undersampling edited nearest neighbors}:
It removes observations whose actual class label differs from the class of at least $\frac{k}{2}$ nearest neighbors. It can undersample indiscriminately over all the existing classes or a subset of the most populated classes.

\item \textit{Oversampling random}:
It
%starting from classes with fewer observations than the most populated class,
randomly duplicates observation of the less populated classes such that all classes have the same population (equal to the initial most populated class).
The main limitation is that the duplicated observations possibly lead to an overfitting problem.

\item \textit{Oversampling synthetic minority over-sampling technique (SMOTE)}:
It randomly selects an observation $a$ from the less populated class and identifies its $k$ nearest neighbors from the same class. 
A synthetic observations $s$ is  generated as a convex combination of $a$ and an observation $b$ randomly selected from its $k$ nearest neighbors.

\item \textit{Oversampling k-means SMOTE}:
It uses the well-known k-means unsupervised learning algorithm to generate new observations of the less populated class in safe and crucial input dataset areas. This technique avoids the generation of noise and effectively overcomes imbalances between and within classes.
% K-means SMOTE works in three steps:
% \begin{itemize}
% \item use k-means on the entire input dataset,
% \item distribute the number of samples to generate across clusters:
% \begin{itemize}
% \item filter clusters having a high number of observations being part of the most populated class,
% \item create and assign synthetic observations to clusters where minority class samples are sparsely distributed, 
% \end{itemize}
% \item oversample each filtered cluster using SMOTE.
% \end{itemize}

\item \textit{Oversampling adaptive synthetic (ADASYN)}:
It uses a weighted distribution for different minority class observations according to their learning phase difficulty. 
Then, less populated classes are enriched by new synthetic observations of such harder to learn observations than those minority examples that are easier to learn.~\cite{he2008adaptive}.
\end{itemize}
%----------------------------------------------------
\subsection{Model training and testing}\label{SEC:MT}
\begin{figure}[!thb]
\centering
\includegraphics[width=0.475\textwidth]{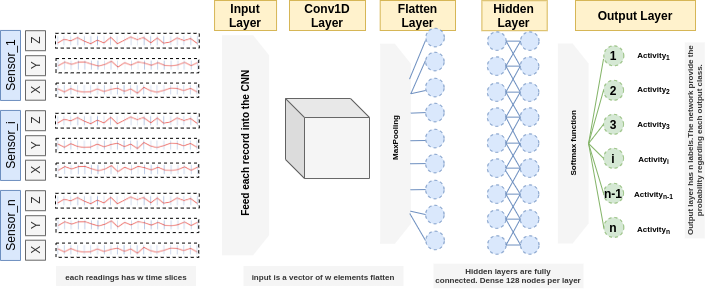}
\caption{Integrated CNN structure.}\label{fig:cnn}
\end{figure}

In such a module, B-HAR, starting from the pre-processed input dataset, takes a known set of input data and responses (training dataset) and trains a model to generate reasonable predictions of new data (testing dataset). 
In particular, B-HAR provides to the user with the possibility of training ten different pattern recognition models: seven machine learning models, i.e., k-Nearest Neighbours (kNN), weighted kNN (wkNN), Linear Discriminant Analysis (LDA), Quadratic Discriminant Analysis (QDA), Support Vector Machine (SVM), Random Forests (RF), Decision Trees (DT), and three deep learning model, i.e., Convolutional Neural Network (CNN), Long Short-Term Memory (LSTM), Recurrent Neural Network (RNN). 
SVM and CNN are the least efficient in terms of training time. 
Instead, LDA and QDA are the least efficient in terms of required memory space.\\
Figure~\ref{fig:cnn} shows the structure of the CNN model defined in B-HAR. It takes in input data from different sensors, providing as an outcome the probabilities of the input data to be recognized as one of the investigated activities.

B-HAR does not perform the usual model training phase, but it applies the grid search training approach, which exhaustively tests candidates from a given grid of parameters. 
For example, the SVM model learns from the training dataset by using different configuration parameters (e.g., kernel function $\in$ [linear, polynomial, sigmoid, radial~basis~function], penalty$\in$ [li,l2], or loss function$\in$ [hinge, squared hinge]) and it is tested on the testing dataset for each configuration. 
On such utilization, B-HAR returns, for each tested model, the configuration that achieved the best results in terms of sensitivity, specificity, precision, f1-score, and accuracy.

However, we want to emphasize that \textbf{B-HAR's primary goal is not to} identify the best recognition model or achieve the best performance results. Instead, it provides a valuable instrument in defining the HAR data processing workflow reusable for their self-designed pattern recognition models. 
For such purposes, besides the integrated models, B-HAR allows users to integrate (aka., plug-in) their self-designed models and test them with the workflow defined with B-HAR.

%----------------------------------------------------
\subsection{Performance measures}
Generally the quality of a pattern recognition model is measured by using accuracy as the main parameter. 
However, this metric could not be precise when an unbalanced training dataset is given in input or when working with multi-class data.
To overcome such an issue, more representative metrics are used that can be extracted from the confusion matrix~\cite{powers2011evaluation}. 
A confusion matrix is a specific item that clearly visualizes the performance of the model.
The mathematical representation of a multi-class confusion matrix is as follows, where rows represent the instances in a predicted class, and columns represent the instances in an actual class.
\begin{align*}
\begin{matrix}
\text{predicted}&&
\end{matrix}\\
C=\text{\parbox[t]{2mm}{\rotatebox[origin=c]{90}{actual}}}~~
\begin{bmatrix}
c_{11} & ... & c_{1n}\\ 
\vdots  & \ddots  & \\ 
c_{n1} &  & c_{nn}
\end{bmatrix}
\end{align*}
The confusion elements for each class are given by:
\begin{itemize}
\item true  positives: $tp = \sum c_{ii}$;
\item false positives: $fp = \sum_{l=1}^n c_{li} - tp_i$;
\item false negatives: $fn = \sum_{l=1}^n c_{il} - tp_i$;
\item true  negatives: $tn = \sum_{l=1}^n \sum_{k=1}^n c_{lk} - tp_i - fp_i - fn_i$.    
\end{itemize}
On the basis of the confusion matrix the following model quality metrics are computed by B-HAR: 

\begin{tabular}{c c c}
\centering
&&\\
~~~~~$P_{recision} = \frac{ tp}{ tp +  fp}$ &  $S_{pecificity} = \frac{ tn}{ fp +  tn}$&\\
&&\\
~~~~~$S_{ensitivity}  = \frac{ tp}{ tp +  fn}$& $A_{ccuracy}  = \frac{ tp +  tn}{ p +  n}$&\\%(4)
&& 
\end{tabular}

\begin{tabular}{ccc}
\centering
~~~~~&$F1_{Score}  =2\times \frac{P_{recision} \times S_{ensitivity}}{P_{recision} + S_{ensitivity}}$&\\
& &
\end{tabular}

Moreover, to better explicate the results and the behavior of the tested models, B-HAR %provides the graph that 
shows the trend of the loss function and the total accuracy at each step of the model training.

Finally, B-HAR provides detailed statistics regarding the execution time of each module.
% !TEX root = ../main.tex
\section{Experimental results}\label{sec:res}
This section presents the results of an extensive experimental campaign performed by using B-HAR on seven of the most popular open-source HAR datasets and two dedicated datasets (i.e., Channel State Information (CSI) and Received Signal Strength Indicator (RSSI) used to recognize the occupancy (empty or occupied) status of an environment starting from the radio signal propagation patterns).
The campaign's goal is twofold: i) testing the flexibility and easiness of use of B-HAR on heterogeneous datasets concerning: the number of subjects, type of activities, and type of sensors, and ii) measuring the performances achieved by machine and deep learning models implemented in B-HAR.
Table~\ref{tab:tested_datasets} presents the main characteristics of the considered datasets. 
Column 1 shows the datasets' names $[$reference$]$. 
Columns 2, 3, and 4 show the number of testing subjects, activities, and used sensors, where A refers to accelerometers and G to gyroscopes.
Column 5 indicates the data sampling frequency, and finally, column 6 presents the dimension of the time window and overlapping fragment in seconds and references the article where such window and overlap dimension were used.
% WISDM v1~\cite{kwapisz2011activity},
% WISDM v2~\cite{lockhart2011design},
% DAPHNET~\cite{bachlin2009wearable},
% HHAR~\cite{stisen2015smart}, 
% PAPAM~\cite{reiss2012introducing},
% mHealth~\cite{banos2014mhealthdroid}
% REALDISP~\cite{banos2012benchmark}
\begin{table}[ht!]
\centering
\resizebox{0.48\textwidth}{!}{
\begin{tabular}{c|cccc|c}
\hline
\hline
\makecell{Dataset\\ name}        &
\makecell{$\#$ of\\ Subjects}    &
\makecell{$\#$ of\\ Activities}  &
%\makecell{$\#$ of\\ Sensors }   &
\makecell{Used\\ Sensors}        &
\makecell{Sampling\\ Frequency}  &
\makecell{(w,o)$[$ref$]$}\\
\hline
\hline
WISDM v1~\cite{kwapisz2011activity}     & 51&  18&  1 (A)    & 20 Hz& (10, 0)\cite{weiss2019smartphone}\\
WISDM v2~\cite{lockhart2011design}      &225&  6&  1 (A)    & 20 Hz& (10, 0)\cite{weiss2019smartphone}\\
%Opportunity~\cite{roggen2010collecting} &  4& 18& 23 (A, G) & & \cite{}(,) \\
DAPHNET~\cite{bachlin2009wearable}      & 10&  2&  3 (A)    &      65 Hz & (3, 1)\cite{demrozi2019towards} \\
HHAR (phone)~\cite{stisen2015smart}     &  9&  6&  2 (A, G)  & 200 Hz & (2, 1)\cite{stisen2015smart}\\
HHAR (watch)~\cite{stisen2015smart}     &  9&  6&  2 (A, G)  &  200 Hz & (2, 1)\cite{stisen2015smart}\\
%Skoda~\cite{zappi2007activity}          &  1& 10& 20 (A)    & & \cite{}(,) \\
PAPAM~\cite{reiss2012introducing}       & 9&  14&  3 (A, G) & 100 Hz& (5, 1)\cite{reiss2012introducing} \\
%UCI-HAR~\cite{anguita2013public}        & 30&  6&  1 (A, G) & & \cite{}(,) \\
%USC-HAD~\cite{zhang2012usc}             & 14& 12&  1 (A)    & & \cite{}(,) \\
%HHAR~\cite{stisen2015smart}             &  9&  6&  2 (A, G) & & \cite{}(,) \\
mHealth~\cite{banos2014mhealthdroid}    & 10& 12&  3 (A, G) & 50 Hz& (5, 2.5)\cite{nguyen2015recognizing} \\

BLE RSSI~\cite{demrozi2021date}   & 4& 2 &  4 (BLE) & 60 Hz& (1, 0)\cite{demrozi2021date}\\
802.11ac CSI~\cite{demrozi2021estimating}    & 3& 2 & 12 (AP)  &  40 Hz&  (1, 0)\cite{demrozi2021estimating}\\
%REALDISP~\cite{banos2012benchmark}      & 17& 30&  9 (A, G) &50 Hz & \cite{wilson2015human} (9, 0) \\

\hline
\hline
\multicolumn{2}{c}{A = Accelerometer} & \multicolumn{2}{c}{G = Gyroscope}& \multicolumn{2}{c}{BLE device}\\
\multicolumn{2}{c}{AP (Access Points)} &\multicolumn{2}{c}{w = window (seconds)} & \multicolumn{2}{c}{o = overlap (seconds)}%&\multicolumn{}{}{}
\end{tabular}
}\vspace{0.1cm}
\caption{Characteristics of datasets.}\label{tab:tested_datasets}
% \vspace{-0.5cm}
\end{table}\\
%\add{Add a sentence why we not used Opportunity or UCI-HAR datasets.}
\begin{table*}[!th]
\centering
\resizebox{\textwidth}{!}{
\begin{tabular}{l|c|cccccc}
\cline{2-8}
\cline{2-8}
\cline{2-8}
&
Dataset &
$kNN$   &
$LDA$   &
$QDA$   &
%$SVM$   &
$RF$    &
$DT$    &
$CNN$   \\
\hline
\hline
\multirow{9}{*}{\rotatebox{90}{Data Segmentation}}&
WISDM v1~\cite{kwapisz2011activity}     &
[98,61,62,61,62]&[95,12,12,12,12]&[95,20,17,20,17]&[99,76,76,76,76]&[97,55,53,55,53]&[96,26,26,26,26] \\
&WISDM v2~\cite{lockhart2011design}&
[94,69,78,69,78]&[64,40,34,40,34]&[86,65,58,65,58]&[96,90,91,90,91]&[92,77,77,77,77]&[60,62,62,66,60] \\
%&Opportunity~\cite{roggen2010collecting}&[]&[]&[]&[]&[]&[]&[] \\
&DAPHNET~\cite{bachlin2009wearable}&
[16,90,87,90,88]&[09,91,83,91,83]&[08,91,82,91,82]&[14,91,91,91,91]&[08,91,83,91,83]&[12,90,87,90,87] \\
&PAPAM~\cite{reiss2012introducing}&
[90,65,66,65,66]&[90,45,45,45,45]&[92,15,19,15,19]&[93,80,83,80,83]&[91,60,60,60,60]&[90,73,76,73,73] \\
%&Skoda~\cite{zappi2007activity}&[]&[]&[]&[]&[]&[]&[] \\
%&UCI-HAR~\cite{anguita2013public}&[]&[]&[]&[]&[]&[]&[] \\
%&USC-HAD~\cite{zhang2012usc}&[]&[]&[]&[]&[]&[]&[] \\
&HHAR (phone)~\cite{stisen2015smart}&
[96,83,85,83,85]&[88,43,45,43,45]&[88,40,50,40,50]&[98,89,89,88,89]&[93,67,66,67,66]&[96,84,84,84,84] \\
&HHAR (watch)~\cite{stisen2015smart}&
[95,78,82,78,82]&[90,54,52,54,52]&[84,26,27,26,27]&[97,85,85,85,85]&[94,69,69,69,69]&[96,83,83,83,83] \\
&mHealth~\cite{banos2014mhealthdroid}&
[81,76,81,76,81]&[75,38,59,38,59]&[09,91,82,91,82]&[73,85,85,85,85]&[74,77,77,77,77]&[65,80,80,80,80] \\
%&REALDISP~\cite{banos2012benchmark}& []&[]&[]&[]&[]&[]&[] \\
% \cline{2-8}
&RSSI~\cite{demrozi2021date}& [91,91,91,91,91]&[91,91,91,91,91]&[91,91,91,91,91]&[91,91,91,91,91]&[91,91,91,91,91]&[90,91,90,91,91]\\
&CSI~\cite{demrozi2021estimating}& [93,93,93,93,93]&[93,93,93,93,93]&[92,92,92,92,92]&[93,93,93,93,93]&[93,93,93,93,93]&[92,92,92,92,92]\\
\hline
\hline
\multirow{9}{*}{\rotatebox{90}{Features Extraction}}&
WISDM v1~\cite{kwapisz2011activity}     &
[97,56,57,56,57]&[97,49,49,49,49]&[96,42,45,42,45]&[99,84,84,84,84]&[98,60,60,60,60]&[95,25,24,25,24] \\
&WISDM v2~\cite{lockhart2011design}&
[98,92,92,92,92]&[96,86,86,86,86]&[82,61,56,61,56]&[97,94,93,94,93]&[96,87,87,87,87]&[59,41,17,41,17] \\
%&Opportunity~\cite{roggen2010collecting}&[]&[]&[]&[]&[]&[]&[] \\
&DAPHNET~\cite{bachlin2009wearable}&
[35,92,90,92,90]&[64,90,92,90,91]&[09,91,82,91,82]&[57,94,93,94,93]&[56,90,91,90,91]&[09,90,82,90,82] \\
&PAPAM~\cite{reiss2012introducing}&
[92,74,76,74,76]&[95,65,67,65,67]&[96,03,18,03,18]&[95,87,88,87,88]&[95,75,75,75,75]&[89,71,66,71,66] \\
%&Skoda~\cite{zappi2007activity}&[]&[]&[]&[]&[]&[]&[] \\
%&UCI-HAR~\cite{anguita2013public}&[]&[]&[]&[]&[]&[]&[] \\
%&USC-HAD~\cite{zhang2012usc}&[]&[]&[]&[]&[]&[]&[] \\
&HHAR (phone)~\cite{stisen2015smart}&
[96,81,82,81,82]&[90,56,56,56,56]&[83,28,28,28,28]&[99,93,93,93,93]&[97,83,83,83,83]&[97,86,86,86,86] \\
&HHAR (watch)~\cite{stisen2015smart}&
[94,69,69,69,69]&[91,54,54,54,54]&[83,26,24,26,24]&[98,89,89,89,89]&[95,76,77,76,77]&[95,77,78,77,78] \\
&mHealth~\cite{banos2014mhealthdroid}&
[92,89,90,89,90]&[98,88,91,88,91]&[30,71,52,71,52]&[96,94,94,94,94]&[82,87,87,87,87]&[64,82,78,82,78] \\
%&REALDISP~\cite{banos2012benchmark}& []&[]&[]&[]&[]&[]&[] \\
% \cline{2-8}
% ~&~&~&~&~&~&~&~\\
% \cline{2-8}
&RSSI~\cite{demrozi2021date}& [95,95,95,95,95]&[95,95,95,95,95]&[95,95,95,95,95]&[95,95,95,95,95]&[95,95,95,95,95]&[94,95,94,95,94]\\
&CSI~\cite{demrozi2021estimating}& [99,99,99,99,99]&[99,99,99,99,99]&[97,98,97,97,97]&[99,99,99,99,99]&[99,99,99,99,99]&[99,99,99,99,99]\\
\hline
\hline
\multicolumn{1}{c}{}&\multicolumn{7}{c}{\textit{Classification results are shown in terms of [\textbf{Specificity, Sensitivity, Precision, Accuracy, F1-Score}]}}
\end{tabular}
}\vspace{0.1cm}
\caption{B-HAR results on segment/features data representation of the most famous HAR datasets.}\label{tab:baseline_results_1}
% \vspace{-0.5cm}
\end{table*}
B-HAR provides the users with two python function calls.
The first takes as input the target dataset and returns statistics that help the user define the B-HAR object configuration.
In particular, such function call returns:
\begin{itemize}
    \item basic knowledge on the presence of NaN or Inf data;
    \item the distribution of observations concerning testers and activities, helping the user in the identification of possible imbalance issues;
    \item the boxplot distribution for all or specific activities or testers, helping the user in the identification of outlier distribution or noisy data.
\end{itemize}

The second function call takes in input the configuration file and it returns the following information:
\begin{itemize}
    \item the performance metrics discussed above;
    \item if using the B-HAR models, the most performing configuration returned by the grid search engine;
    \item the plot of the loss and accuracy variation during the training process;
    \item if used, the most important features selected by the feature selection model;
    \item if required, the segmented or features dataset;
    \item if required, execution time statistics for each computation module and sub-modules. 
\end{itemize}

Concerning the experimental analysis, Table~\ref{tab:baseline_results_1} shows the results achieved by using B-HAR on the datasets of Table~\ref{tab:tested_datasets} by applying six of the ten pattern recognition models reported in Section~\ref{SEC:MT}.
The configuration of B-HAR used for the experiments is shown by the gray rows of Table~\ref{tab:config_parameters}. In the first block of Table~\ref{tab:baseline_results_1}, data segmentation has been applied. Instead, in the second block of Table~\ref{tab:baseline_results_1}, feature extraction in the time and frequency domains has been used.
%\add{Add a sentence showing why such tests are necessary to the others.}

By observing the results reported in Table~\ref{tab:baseline_results_1}, the user can identify which model performs better for each metric at varying of the considered datasets. 
Of course, the users can provide B-HAR with new datasets and HAR models, configuring the framework by following different workflow.\\
In this way, B-HAR is proposed as a baseline framework for fairly comparing existing and new HAR approaches and easily identifying the one that provides the most accurate results, given the target dataset.\\

\section{HAR design workflow issues}\label{sec:dis}
% we discuss and analyze different aspects of the HAR workflow since many methodologies published in the existing literature are complex to be compared and do not provide precise information regarding the processing steps.  
In this article, we presented an overview of the fundamental processing steps with a primary focus on the definition of the correct order in which such steps have to be applied (Figure~\ref{fig:b-har_structure}) since many methodologies published in the existing literature are complex to be replicated and do not provide precise information regarding the processing steps. 
In the following, we discuss the essential information that HAR methodologies need to provide since comparing solutions with different experimental setups is challenging because accuracy and other factors depend on the performed experimental processing steps.

% \subsection{Dataset information's}
Generally, detailed information related to the collected data and setting up the experimental setup are not provided. 
However, a complete description of the collected data is of primary importance. 
The following information concerning the input dataset needs to be provided:
\begin{itemize}
\item number of testing subjects, gender, age, hights, weight;
\item existing pathologies, left-handed or right-handed;
\item the number of used data collection devices, precise position, and axis heading of each device. 
\end{itemize}
Furthermore, the information regarding the methods of carrying out the data collection phases is fundamental for a correct analysis of the data, for example:
\begin{itemize}
\item the environment (e.g., indoor, outdoor and dimension);
\item the performed activities;
\item how the subjects are trained to carrying out the activities?;
\item data are collected as one activity for each session or many activities for each session?.
\end{itemize}
This last point is crucial since the datasets present many activities for each session, so the transition between activities needs to be handled.

% \subsection{Processing steps}
Beyond the issues concerning the data and the data collection approach, a significant issue in the proposed methodologies concerns the missing details. 
Table~\ref{tab:rec} provides 13 basic recommendations related to the essential information that authors need to provide when proposing a HAR methodology.
Column 2 presents the item of interest. Instead, Column 3 presents the recommendations related to the item. 
\begin{table}[!htb]
\centering
\begin{tabular}{c|c|c}
\cline{2-3}
            &   Info. related to    &   Info. to provide    \\
\hline
\hline

R1 & \makecell{Data source\\ devices\footnote{e.g., smartphone, smartwatch, single board computers (SBC) or standalone devices}} &    
\makecell{manufacturer, model,\\ cost, SO/firmware version}\\
\hline

R2 & \makecell{Sensor\\model}            &  
\makecell{manufacturer, model,\\ sampling frequency} \\
\hline \hline

R3 & \makecell{Dataset\\information's}   &    
\makecell{\# of samples per:\\ subject, activities\\ and sessions}\\
\hline

R4 & \makecell{Data\\errors}             &    
\makecell{used handling\\ errors technique}\\
\hline

R5 & Noise                               &
\makecell{detailed filtering\\ information's} \\
\hline\hline

R6 & Segmentation                        &    
\makecell{window/overlap size and\\discussion concerning such choice}\\
\hline

R7 & \makecell{Feature\\ extraction}     &    
\makecell{features domain and reference\\ to the used features}\\
\hline\hline

R8 & \makecell{Train/Test\\approach}     &    
\makecell{inter-subject, intra-subject/session\\and precise split parameters\\(i.e., \# of subjects, \# of\\ samples per subject and class)}\\
\hline

R9 & Normalization                       &    
\makecell{definition of the used technique\\ and discussion concerning such choice} \\
\hline

R10& \makecell{Feature\\selection}       &    
\makecell{definition of the used technique\\ and discussion concerning such choice} \\
\hline

R11& Balancing                           &    
\makecell{definition of the used technique\\ and discussion concerning such choice} \\
\hline\hline

R12& \makecell{HAR\\model}               &    
\makecell{precise description concerning\\tested model and its configuration} \\
\hline\hline

R13& \makecell{Performance\\metrics}     &
\makecell{definition of the performance metrics\\ and discussion of the achieved results} \\
\hline
\hline
\end{tabular}
\vspace{0.1cm}
\caption{Recommendations for HAR methodologies characteristics exploitation's.}\label{tab:rec}
\end{table}

% Therefore, the exploitation of the introduced recommendations provides richer context and analysis information to adapt to more complex human actions.
B-HAR aim is not to standardize the HAR workflow since the performance of the generated model is strongly related to the data source (i.e., R1 to R3), the single processing steps (i.e., R4 to R11), and the configuration of the HAR model (i.e., R12). 
As for the performance measures, it is widely acknowledged that new performance measures metrics (e.g., robustness against adversarial attacks or true generalizability across datasets) that measure the generalization capabilities of the models are needed. 
Based on such recommendations, B-HAR provides the capability to reproduce the analysis workflow easily and minimizes the users' errors due to the application of the processing steps in the wrong order. 
%\input{sec/rel.tex}
% !TEX root = ../main.tex
\section{Conclusions and future extensions}\label{sec:conc}
HAR based on time series data includes several processing steps that affect the quality of the achieved results.  
However, no clear pipeline illustrates how such steps have to be applied. 
Thus, researchers might make elementary mistakes that affect the quality of the activity recognition model.
This article proposed B-HAR, a framework that facilitates the study of the behavior of the most famous pattern recognition models in the context of HAR.
B-HAR allows the users to define elaboration pipelines, including different pre-processing steps, such as noise removal, segmentation, feature extraction, normalization, feature selection, and balance, in the correct order. 
Moreover, users can utilize B-HAR to test self-designed HAR models by reusing already designed HAR workflows. 
% it returns the evaluation metrics for eight state-of-the-art recognition models and provides the possibility to integrate new recognition models different from those already present in B-HAR.
This reduces the possibility of the incorrect design of HAR pipelines and provides the users with a baseline framework for fairly comparing the corresponding results.

Future work will regard creating a GUI, the extension of each sub-module with a more extensive set of state-of-the-art data processing techniques, integration of new recognition,  regression, transfer learning models, and implementation of new performance metrics. 

\ifCLASSOPTIONcaptionsoff
  \newpage
\fi

\bibliographystyle{IEEEtran}
% \bibliography{IEEEabrv,biblio}{}

\begin{thebibliography}{10}
\providecommand{\url}[1]{#1}
\csname url@samestyle\endcsname
\providecommand{\newblock}{\relax}
\providecommand{\bibinfo}[2]{#2}
\providecommand{\BIBentrySTDinterwordspacing}{\spaceskip=0pt\relax}
\providecommand{\BIBentryALTinterwordstretchfactor}{4}
\providecommand{\BIBentryALTinterwordspacing}{\spaceskip=\fontdimen2\font plus
\BIBentryALTinterwordstretchfactor\fontdimen3\font minus
  \fontdimen4\font\relax}
\providecommand{\BIBforeignlanguage}[2]{{%
\expandafter\ifx\csname l@#1\endcsname\relax
\typeout{** WARNING: IEEEtran.bst: No hyphenation pattern has been}%
\typeout{** loaded for the language `#1'. Using the pattern for}%
\typeout{** the default language instead.}%
\else
\language=\csname l@#1\endcsname
\fi
#2}}
\providecommand{\BIBdecl}{\relax}
\BIBdecl

\bibitem{bianchi2019iot}
V.~Bianchi, M.~Bassoli, G.~Lombardo, P.~Fornacciari, M.~Mordonini, and
  I.~De~Munari, ``Iot wearable sensor and deep learning: An integrated approach
  for personalized human activity recognition in a smart home environment,''
  \emph{IEEE Internet of Things Journal}, vol.~6, no.~5, pp. 8553--8562, 2019.

\bibitem{poli2020impact}
A.~Poli, G.~Cosoli, L.~Scalise, and S.~Spinsante, ``Impact of wearable
  measurement properties and data quality on adls classification accuracy,''
  \emph{IEEE Sensors Journal}, 2020.

\bibitem{demrozi2021exploiting}
F.~Demrozi, N.~Serlonghi, C.~Turetta, C.~Pravadelli, and G.~Pravadelli,
  ``Exploiting bluetooth low energy smart tags for virtual coaching,'' in
  \emph{2021 IEEE 7th World Forum on Internet of Things (WF-IoT)}.\hskip 1em
  plus 0.5em minus 0.4em\relax IEEE, 2021, pp. 470--475.

\bibitem{UN19}
{Department of Economic and Social Affairs, United Nations}, \emph{World
  Population Prospects 2019: Highlights}.\hskip 1em plus 0.5em minus
  0.4em\relax United Nations, 2019.

\bibitem{demrozi2019towards}
F.~Demrozi, R.~Bacchin, S.~Tamburin, M.~Cristani, and G.~Pravadelli, ``Towards
  a wearable system for predicting the freezing of gait in people affected by
  parkinson's disease,'' \emph{IEEE journal of biomedical and health
  informatics}, 2019.

\bibitem{demrozi2019indoor}
F.~Demrozi, V.~Bragoi, F.~Tramarin, and G.~Pravadelli, ``An indoor localization
  system to detect areas causing the freezing of gait in parkinsonians,'' in
  \emph{2019 Design, Automation \& Test in Europe Conference \& Exhibition
  (DATE)}.\hskip 1em plus 0.5em minus 0.4em\relax IEEE, 2019, pp. 952--955.

\bibitem{lara2012survey}
O.~D. Lara and M.~A. Labrador, ``A survey on human activity recognition using
  wearable sensors,'' \emph{IEEE communications surveys \& tutorials}, vol.~15,
  no.~3, pp. 1192--1209, 2012.

\bibitem{fullerton2017recognizing}
E.~Fullerton, B.~Heller, and M.~Munoz-Organero, ``Recognizing human activity in
  free-living using multiple body-worn accelerometers,'' \emph{IEEE Sensors
  Journal}, vol.~17, no.~16, pp. 5290--5297, 2017.

\bibitem{demrozi2021inertial}
F.~Demrozi, M.~Jereghi, and G.~Pravadelli, ``Towards the automatic data
  annotation for human activity recognition based on wearables and ble
  beacons,'' in \emph{2021 IEEE International Symposium on Inertial Sensors and
  Systems (INERTIAL)}.\hskip 1em plus 0.5em minus 0.4em\relax IEEE, 2021, pp.
  1--4.

\bibitem{pucci2020human}
L.~Pucci, E.~Testi, E.~Favarelli, and A.~Giorgetti, ``Human activities
  classification using biaxial seismic sensors,'' \emph{IEEE Sensors Letters},
  vol.~4, no.~10, pp. 1--4, 2020.

\bibitem{gambi2020adl}
E.~Gambi, G.~Temperini, R.~Galassi, L.~Senigagliesi, and A.~De~Santis, ``Adl
  recognition through machine learning algorithms on iot air quality sensor
  dataset,'' \emph{IEEE Sensors Journal}, vol.~20, no.~22, pp.
  13\,562--13\,570, 2020.

\bibitem{lu2020efficient}
J.~Lu, X.~Zheng, M.~Sheng, J.~Jin, and S.~Yu, ``Efficient human activity
  recognition using a single wearable sensor,'' \emph{IEEE Internet of Things
  Journal}, 2020.

\bibitem{donahue2015long}
J.~Donahue, L.~Anne~Hendricks, S.~Guadarrama, M.~Rohrbach, S.~Venugopalan,
  K.~Saenko, and T.~Darrell, ``Long-term recurrent convolutional networks for
  visual recognition and description,'' in \emph{Proceedings of the IEEE
  conference on computer vision and pattern recognition}, 2015, pp. 2625--2634.

\bibitem{burgos2020ear}
C.~P. Burgos, L.~G{\"a}rtner, M.~A.~G. Ballester, J.~Noailly, F.~St{\"o}cker,
  M.~Sch{\"o}nfelder, T.~Adams, and S.~Tassani, ``In-ear accelerometer-based
  sensor for gait classification,'' \emph{IEEE Sensors Journal}, vol.~20,
  no.~21, pp. 12\,895--12\,902, 2020.

\bibitem{demrozi2020human}
F.~{Demrozi}, G.~{Pravadelli}, A.~{Bihorac}, and P.~{Rashidi}, ``Human activity
  recognition using inertial, physiological and environmental sensors: A
  comprehensive survey,'' \emph{IEEE Access}, pp. 1--1, 2020.

\bibitem{demrozi2020pain}
F.~{Demrozi}, G.~{Pravadelli}, P.~J. {Tighe}, A.~{Bihorac}, and P.~{Rashidi},
  ``Joint distribution and transitions of pain and activity in critically ill
  patients,'' in \emph{2020 42nd Annual International Conference of the IEEE
  Engineering in Medicine Biology Society (EMBC)}, 2020, pp. 4534--4538.

\bibitem{black2004pervasive}
J.~Black, W.~Segmuller, N.~Cohen, B.~Leiba, A.~Misra, M.~Ebling, and E.~Stern,
  ``Pervasive computing in health care: Smart spaces and enterprise information
  systems,'' in \emph{MobiSys 2004 Workshop on Context Awareness}, 2004.

\bibitem{kameas2010pervasive}
A.~Kameas and I.~Calemis, ``Pervasive systems in health care,'' in
  \emph{Handbook of ambient intelligence and smart environments}.\hskip 1em
  plus 0.5em minus 0.4em\relax Springer, 2010, pp. 315--346.

\bibitem{kanjo2019deep}
E.~Kanjo, E.~M. Younis, and C.~S. Ang, ``Deep learning analysis of mobile
  physiological, environmental and location sensor data for emotion
  detection,'' \emph{Information Fusion}, vol.~49, pp. 46--56, 2019.

\bibitem{brownlee2016grid}
J.~Brownlee, ``How to grid search hyperparameters for deep learning models in
  python with keras,'' \emph{Retrieved April}, vol.~20, p. 2018, 2016.

\bibitem{kwapisz2011activity}
J.~R. Kwapisz, G.~M. Weiss, and S.~A. Moore, ``Activity recognition using cell
  phone accelerometers,'' \emph{ACM SigKDD Explorations Newsletter}, vol.~12,
  no.~2, pp. 74--82, 2011.

\bibitem{roggen2010collecting}
D.~Roggen, A.~Calatroni, M.~Rossi, T.~Holleczek, K.~F{\"o}rster,
  G.~Tr{\"o}ster, P.~Lukowicz, D.~Bannach, G.~Pirkl, A.~Ferscha \emph{et~al.},
  ``Collecting complex activity datasets in highly rich networked sensor
  environments,'' in \emph{2010 Seventh international conference on networked
  sensing systems (INSS)}.\hskip 1em plus 0.5em minus 0.4em\relax IEEE, 2010,
  pp. 233--240.

\bibitem{bachlin2009wearable}
M.~Bachlin, M.~Plotnik, D.~Roggen, I.~Maidan, J.~M. Hausdorff, N.~Giladi, and
  G.~Troster, ``Wearable assistant for parkinson’s disease patients with the
  freezing of gait symptom,'' \emph{IEEE Transactions on Information Technology
  in Biomedicine}, vol.~14, no.~2, pp. 436--446, 2009.

\bibitem{reiss2012introducing}
A.~Reiss and D.~Stricker, ``Introducing a new benchmarked dataset for activity
  monitoring,'' in \emph{2012 16th International Symposium on Wearable
  Computers}.\hskip 1em plus 0.5em minus 0.4em\relax IEEE, 2012, pp. 108--109.

\bibitem{zappi2007activity}
P.~Zappi, T.~Stiefmeier, E.~Farella, D.~Roggen, L.~Benini, and G.~Troster,
  ``Activity recognition from on-body sensors by classifier fusion: sensor
  scalability and robustness,'' in \emph{2007 3rd international conference on
  intelligent sensors, sensor networks and information}.\hskip 1em plus 0.5em
  minus 0.4em\relax IEEE, 2007, pp. 281--286.

\bibitem{anguita2013public}
D.~Anguita, A.~Ghio, L.~Oneto, X.~Parra, and J.~L. Reyes-Ortiz, ``A public
  domain dataset for human activity recognition using smartphones.'' in
  \emph{Esann}, 2013.

\bibitem{zhang2012usc}
M.~Zhang and A.~A. Sawchuk, ``Usc-had: a daily activity dataset for ubiquitous
  activity recognition using wearable sensors,'' in \emph{Proceedings of the
  2012 ACM Conference on Ubiquitous Computing}.\hskip 1em plus 0.5em minus
  0.4em\relax ACM, 2012, pp. 1036--1043.

\bibitem{stisen2015smart}
A.~Stisen, H.~Blunck, S.~Bhattacharya, T.~S. Prentow, M.~B. Kj{\ae}rgaard,
  A.~Dey, T.~Sonne, and M.~M. Jensen, ``Smart devices are different: Assessing
  and mitigatingmobile sensing heterogeneities for activity recognition,'' in
  \emph{Proceedings of the 13th ACM conference on embedded networked sensor
  systems}, 2015, pp. 127--140.

\bibitem{banos2014mhealthdroid}
O.~Banos, R.~Garcia, J.~A. Holgado-Terriza, M.~Damas, H.~Pomares, I.~Rojas,
  A.~Saez, and C.~Villalonga, ``mhealthdroid: a novel framework for agile
  development of mobile health applications,'' in \emph{International workshop
  on ambient assisted living}.\hskip 1em plus 0.5em minus 0.4em\relax Springer,
  2014, pp. 91--98.

\bibitem{banos2012benchmark}
O.~Ba{\~n}os, M.~Damas, H.~Pomares, I.~Rojas, M.~A. T{\'o}th, and O.~Amft, ``A
  benchmark dataset to evaluate sensor displacement in activity recognition,''
  in \emph{Proceedings of the 2012 ACM Conference on Ubiquitous
  Computing}.\hskip 1em plus 0.5em minus 0.4em\relax ACM, 2012, pp. 1026--1035.

\bibitem{lockhart2011design}
J.~W. Lockhart, G.~M. Weiss, J.~C. Xue, S.~T. Gallagher, A.~B. Grosner, and
  T.~T. Pulickal, ``Design considerations for the wisdm smart phone-based
  sensor mining architecture,'' in \emph{Proceedings of the Fifth International
  Workshop on Knowledge Discovery from Sensor Data}, 2011, pp. 25--33.

\bibitem{pratama2016review}
I.~Pratama, A.~E. Permanasari, I.~Ardiyanto, and R.~Indrayani, ``A review of
  missing values handling methods on time-series data,'' in \emph{2016
  International Conference on Information Technology Systems and Innovation
  (ICITSI)}.\hskip 1em plus 0.5em minus 0.4em\relax IEEE, 2016, pp. 1--6.

\bibitem{sensormotion}
``Sensormotion,'' \url{https://pypi.org/project/sensormotion/}, accessed:
  2020-11-24.

\bibitem{barandas2020tsfel}
M.~Barandas, D.~Folgado, L.~Fernandes, S.~Santos, M.~Abreu, P.~Bota, H.~Liu,
  T.~Schultz, and H.~Gamboa, ``Tsfel: Time series feature extraction library,''
  \emph{SoftwareX}, vol.~11, p. 100456, 2020.

\bibitem{shoaib2015survey}
M.~Shoaib, S.~Bosch, O.~D. Incel, H.~Scholten, and P.~J. Havinga, ``A survey of
  online activity recognition using mobile phones,'' \emph{Sensors}, vol.~15,
  no.~1, pp. 2059--2085, 2015.

\bibitem{bishop2006pattern}
C.~M. Bishop, \emph{Pattern recognition and machine learning}.\hskip 1em plus
  0.5em minus 0.4em\relax springer, 2006.

\bibitem{chandrashekar2014survey}
G.~Chandrashekar and F.~Sahin, ``A survey on feature selection methods,''
  \emph{Computers \& Electrical Engineering}, vol.~40, no.~1, pp. 16--28, 2014.

\bibitem{imbalance}
\BIBentryALTinterwordspacing
G.~Lema{{\^i}}tre, F.~Nogueira, and C.~K. Aridas, ``Imbalanced-learn: A python
  toolbox to tackle the curse of imbalanced datasets in machine learning,''
  \emph{Journal of Machine Learning Research}, vol.~18, no.~17, pp. 1--5, 2017.
  [Online]. Available: \url{http://jmlr.org/papers/v18/16-365}
\BIBentrySTDinterwordspacing

\bibitem{he2008adaptive}
H.~He, Y.~Bai, E.~Garcia, and S.~A. Li, ``Adaptive synthetic sampling approach
  for imbalanced learning. ieee international joint conference on neural
  networks. 2008,'' 2008.

\bibitem{powers2011evaluation}
D.~M. Powers, ``Evaluation: from precision, recall and f-measure to roc,
  informedness, markedness and correlation,'' \emph{--}, 2011.

\bibitem{weiss2019smartphone}
G.~M. Weiss, K.~Yoneda, and T.~Hayajneh, ``Smartphone and smartwatch-based
  biometrics using activities of daily living,'' \emph{IEEE Access}, vol.~7,
  pp. 133\,190--133\,202, 2019.

\bibitem{nguyen2015recognizing}
L.~T. Nguyen, M.~Zeng, P.~Tague, and J.~Zhang, ``Recognizing new activities
  with limited training data,'' in \emph{Proceedings of the 2015 ACM
  International Symposium on Wearable Computers}, 2015, pp. 67--74.

\bibitem{demrozi2021date}
F.~Demrozi, F.~Chiarani, and G.~Pravadelli, ``A low-cost ble-based distance
  estimation, occupancy detection and counting system,'' in \emph{ACM/IEEE
  DATE}, 2021.

\bibitem{demrozi2021estimating}
F.~Demrozi, C.~Turetta, F.~Chiarani, P.~H. Kindt, and G.~Pravadelli,
  ``Estimating indoor occupancy through low-cost ble devices,'' \emph{IEEE
  Sensors Journal}, 2021.

\end{thebibliography}
% Generated by IEEEtran.bst, version: 1.14 (2015/08/26)

\begin{IEEEbiography}
[{\includegraphics[width=0.9in,height=1.25in,clip,keepaspectratio]{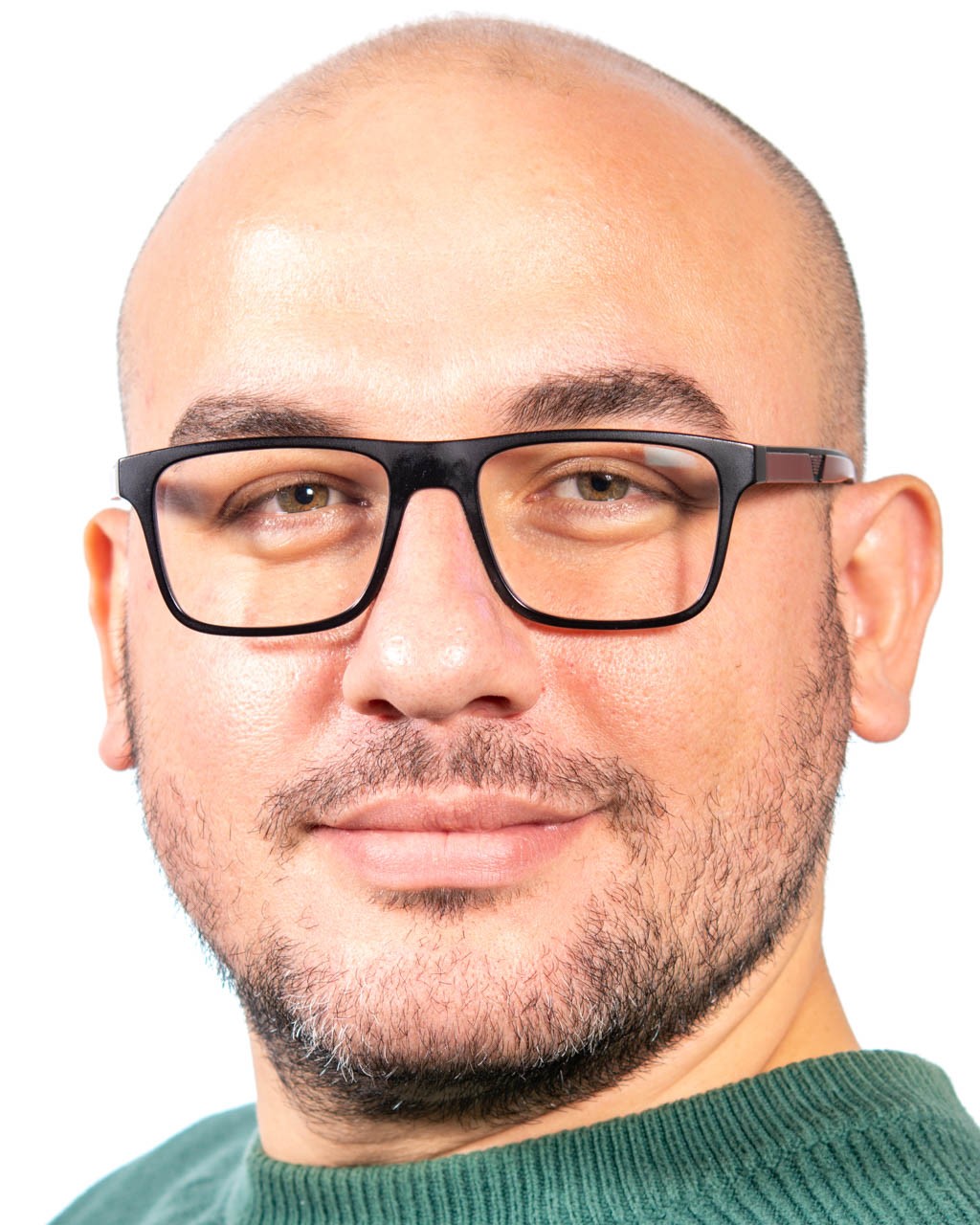}}]{\textbf{Florenc Demrozi,}} PhD in computer science, IEEE member, received the B.S. and M.E. degrees in Computer Science and Engineering from the University of Verona, Italy, respectively in 2014 and 2016, and the Ph.D. degree in Computer Science from University of Verona, Italy, in 2020. He is currently an Associate Professor in Biomedical Engineering at the Department of Electrical Engineering and Computer Science, University of Stavanger, Norway, working on Human Activity Recognition (HAR), Ambient Assisted Living (AAL), Internet of Medical Things (IoMT), Sensors and Measurements. 
\end{IEEEbiography}
% \vspace{-1.25cm}
\begin{IEEEbiography}
[{\includegraphics[width=1in,height=1.1in,clip]{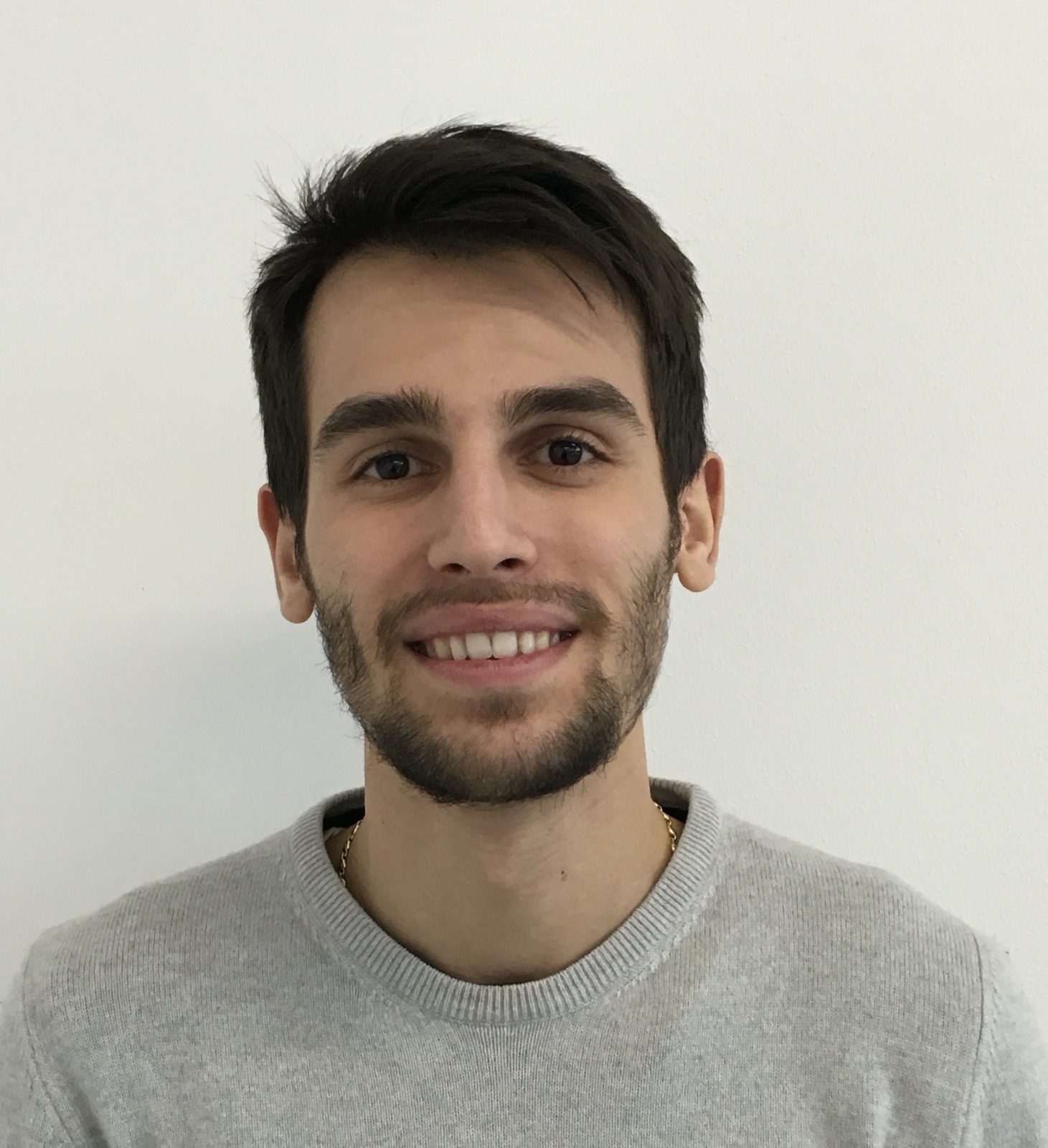}}]{\textbf{Cristian Turetta,}} received the B.S. and M.E. degrees in Computer Science and Engineering from the University of Verona, Italy, respectively in 2017 and 2020. He is currently a research fellow at the Department of Computer Science, University of Verona, Italy working on Ambient Assisted Living (AAL), Internet of Things (IoT) and IoT Security.
\end{IEEEbiography}
% \vspace{-1.5cm}
\begin{IEEEbiography}
[{\includegraphics[width=0.9in,height=1.25in,clip,keepaspectratio]{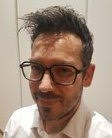}}]{\textbf{Graziano Pravadelli,}} PhD in computer science, IEEE senior member, IFIP 10.5 WG member, is full professor of information processing systems at the Computer Science Department of the University of Verona (Italy) since 2018. In 2007 he cofounded EDALab s.r.l., an SME working on the design of IoT-based monitoring systems. His main interests focus on system-level modeling, simulation and semi-formal verification of embedded systems, as well as on their application to develop IoT-based virtual coaching platforms for people with special needs. In the previous contexts, he collaborated in several national and European projects and he published more than 120 papers in international conferences
and journals.
\end{IEEEbiography}

\end{document}